\documentclass[fleqn,usenatbib]{mnras}

\usepackage{graphicx}   
\usepackage{amsmath}    
\usepackage{amssymb}    
\usepackage{txfonts}

\newcommand{\bc}{\begin{center}}
\newcommand{\ec}{\end{center}}
\usepackage{color}

\title[Comparing Galaxy morphology]{Comparing galaxy morphology in hydrodynamical simulation and in semi-analytic model}

\author[Wang et al. ]
       {Lan Wang$^{1}$\thanks{Email:
           wanglan@bao.ac.cn}, Dandan Xu$^{2,3}$, Liang Gao$^{1,4}$, Qi Guo$^{1,4}$, Yan Qu$^{1}$, Jun Pan$^{1}$
        \\      
        $^1$Key Laboratory for Computational Astrophysics, National
Astronomical Observatory, Chinese Academy of Sciences,\\
Datun Road 20A, Beijing 100012, China\\
        $^2$Institute for Advanced Studies, Tsinghua University, Beijing, 100084, China \\
        $^3$Tsinghua Centre for Astrophysics, Tsinghua University, Beijing, 100084, China \\
        $^4$School of Astronomy and Space Science, University of Chinese Academy of Sciences
	}
 
\date{Accepted 2018 ???? ??. 
      Received 2018 ???? ??; 
      in original form 2018 ???? ??}
\pubyear{2018}

\begin{document}

\label{firstpage}

\pagerange{\pageref{firstpage}--\pageref{lastpage}} 
\maketitle


\begin{abstract}
We compare galaxy morphology predicted by the Illustris hydrodynamical simulation and a semi-analytic model (SAM) grafted in the halo merger trees from the Illustris-Dark Matter simulation. Morphology is classified according to the luminous profile and the kinematic bulge-to-total ratio for Illustris galaxy, and the bulge-to-total stellar mass ratio for SAM galaxy. For late-type galaxies in the Illustris catalogue, most of their counterparts in the SAM model have the same type, and the consistency between two models is higher for lower mass galaxies. For early-type galaxies in Illustris, the consistency is quite low for the counterparts except for most massive galaxies. By comparing in detail the growth histories of some matched galaxy pairs of Milky Way-mass in Illustris and the SAM model, we notice two aspects of differences in determining galaxy morphology between the two galaxy formation implementations. Firstly, in the SAM, major merger and frequent minor mergers result in the growth of bulges and turn the galaxy into early type, while bulge formation is not connected to mergers as tightly as in SAM for the Illustris galaxies. In addition, the satellite stellar mass can decrease significantly due to tidal stripping before merging into the central galaxy in Illustris, while it does not decrease in the SAM model. This results in less mergers with large (stellar) mass ratios in the Illustris simulation, and less effect of mergers on shaping galaxy morphology.

\end{abstract}


\section{Introduction}
\label{sec:intro}

In the current standard galaxy formation scenario, galaxies reside and form in dark matter haloes\citep{white1978}. The formation and evolution of galaxies involves complicated physical processes, including gas cooling, star formation and stellar feedback, mergers, black hole growth and active galactic nuclei(AGN) feedback\citep{white1991,croton2006, dayal2018}. The degeneracy of these physical processes in determining the final galaxy properties makes it hard to constrain the contribution from each process directly and independently. To study and model the overall galaxy formation and evolution in a cosmological context following these detailed physical processes, two main methods have been employed: semi-analytic galaxy formation models which model luminous galaxies based on dark matter merging tree by implementing empirical formulas describing the physical processes of baryons \citep{white1991,baugh2006}, and the hydrodynamical simulations that solve the dynamics of gas and stellar particles directly \citep{katz1996,springel2005}.

Hydrodynamical simulations have the advantage to be able to study the physical processes in more detail than the semi-analytical method, while the latter requires less computational cost and therefore is more flexible in tuning and testing model parameters. Nevertheless, both approaches are able to reproduce many statistics of galaxy properties at low and high redshifts. While it has been difficult for a long time for hydrodynamical simulations to reproduce galaxy morphology, especially the disk fraction and disk sizes of real galaxies, the most recent improvements have been achieved to be able to produce Milky Way disks with reasonable size with hydrodynamical simulations \citep{vogelsberger2014a, schaye2015}. Disk sizes are also studied in more detail recently in semi-analytic models \citep{guo2011,GP2014}. The most recent comparisons indicate that galaxies have different stellar mass-size relations in the two models \citep{guo2016,mitchell2017}.

It is interesting to see whether the two methods can give in general consistent predictions on galaxy morphology such as disk/elliptical fractions. In semi-analytic models, since baryons are assumed to follow dark matter tightly, for a galaxy, once its morphology is initialized according to the properties of its progenitor halo at early time, its following evolution is closely related to the assembly history of its host dark matter halo. For example, mergers play an important role in transforming disk galaxies into bulge-dominant ones, with major mergers turning galaxies into ellipticals and minor mergers responsible for the growth of bulge component. In hydrodynamical simulations, on the other hand, the morphology of a galaxy is a consequence of dynamical processes of internal star particles, which is not necessarily related to the assembly of the host dark matter halo. In this study, we use the state-of-art hydrodynamical simulation Illustris \citep{vogelsberger2014a,vogelsberger2014b, genel2014, sijacki2015,nelson2015}, and a semi-analytic model result to compare galaxy morphology predicted by the two models. The latter is produced by applying \citet{guo2011} semi-analytic model to the Illustris Dark matter simulation, which is a dark matter-only simulation with exactly the same initial conditions and cosmological parameter set as those of Illustris hydrosimulation. By doing this, we are able to match galaxies one-to-one in the two models and compare their morphologies case by case, to investigate whether the two models are consistent in producing galaxy morphology. 


This paper is organised as follows. Section 2 introduces the simulations and semi-analytic model used in this study, and how we determine morphology type of galaxies in the two methods. In section 3 we first introduce how we match galaxies between the hydro-simulation and semi-analytic results, then we compare the morphologies of the matched galaxies predicted by two models statistically. Section 4 presents a specific example of galaxy merger trees, to compare morphology evolution in the two models in detail. Discussions and conclusions are presented in section 5.

\section{morphology determination in Illustris and the SAM model}
\label{sec:sim}

\subsection{Illustris and Illustris-Dark simulations}

The Illustris project \citep{vogelsberger2014a} comprises a series of cosmological hydrodynamical simulations of galaxy formation, with a volume of $(106.5 \rm Mpc)^3$ but with different mass resolutions. For the highest-resolution run used in this study, it evolves $1820^3$ dark matter and gas particles, with the mass resolution of dark matter particle to be $6.26 \times 10^6 M_{\odot}$, and the initial baryonic mass resolution of $1.26 \times 10^6 \rm M_{\odot}$, respectively. Gravitational dynamics is resolved down to a physical scale of $710$ pc. The Illustris simulations take into account various baryonic processes including gas cooling, stellar evolution and feedback, chemical enrichment, black hole growth and feedback from active galactic nuclei, and successfully reproduce many fundamental properties of observed galaxies, including producing various galaxy morphologies, with consistent stellar and baryonic Tully - Fisher relation as observed \citep{vogelsberger2014a}.

Each Illustris hydrodynamical simulation has a matched dark matter-only run performed  with the same set of parameters and initial conditions. For the matched dark matter-only simulation, it evolves $1820^3$ dark matter particles with a mass resolution of $7.5 \times 10^6 \rm M_{\odot}$. Hereafter, we refer the hydrodynamical run we use as Illustris, and the pure dark matter one as Illustris-Dark.  

For both hydro- and dark matter-only simulations, a friends-of-friends (FOF) algorithm \citep{davis1985} is used to identify dark matter haloes with a minimum particle number of 32. Subhaloes are then identified using the SUBFIND algorithm\citep{springel2001, dolag2009}. For the hydro-simulation, other types of particles such as stellar, gas and black hole particles are attached to the FOF haloes in a secondary linking stage \citep{dolag2009}, and galaxies are associated with subhaloes.

Based on the subhalo catalogue, L-HaloTree \citep[as used for the Millennium simulation,][]{springel2005b} and SubLink \citep{rodriguez2015} codes are applied to the Illustris-dark and the Illustris hydro simulation to construct merging trees.

\subsection{Morphology determination of the Illustris galaxies}

There are two main methods to classify morphologies of galaxies in observation and in simulations. One is based on the luminous profile of a galaxy and the other relies on its kinematic feature. In this paper, we use both methods to define morphology types for the Illustris galaxies. Below we give a brief summary of each method. A detailed comparison between the outcomes of the two methods is presented in section 3. 

The luminous profiles of Illustris galaxies have been studied in  \citet{xu2017}. In their work, the morphology type of a galaxy in a given projection is decided according to the radial surface brightness distribution in the projection. If it is better fitted by a de Voucauleurs profile, the galaxy is classified as an early-type (ET) galaxy; if it is better fitted by an exponential profile, the galaxy is classified as a late-type (LT) galaxy in that projection. For a given galaxy, the type decisions are made in all three different projections along the principal axes of the simulation box. The final light-based morphology type of the galaxy is determined to be the decision that are reached in $\ge 2$ projections.

An alternative definition of morphology type is based on kinematics of galaxies. For Illustris galaxies, \citet{genel2015} provided the kinematic bulge-to-total ratios $(B/T)_{\rm kin}$, defined by two times the fractional mass of stars with circularity parameter less than 0, which is commonly used in the literature to define the bulge fraction of a galaxy \citep{binney1987}. Following the divider adopted by \citet{teklu2015}, in this study, we define galaxies with $(B/T)_{\rm kin}<0.6$ to be kinematics-based late-type and those with $(B/T)_{\rm kin}>0.6$ to be kinematics-based early-type.

\subsection{Semi-analytic model and morphology determination}
In this study, we apply the semi-analytic model of \citet{guo2011} to the dark matter merger tree constructed using L-HaloTree \citep{springel2005b} of Illustris-Dark, to predict galaxy morphology.  The model of \citet{guo2011} is based on a series of semi-analytic models developed by the Munich group \citep{kauffmann1993,croton2006,DLB2007} with substantial updates and additions on the modeling of supernovae feedback and sizes of galactic disks. In \citet{guo2011}, three modes of bulge growth are included: major mergers, minor mergers and disc instability.  After a major merger, all the existent and the newly formed stars are assumed to end up in a spheroidal component, forming an elliptical galaxy.  When a minor merger occurs, a bulge grows by acquiring all the pre-existing stars from the minor progenitor, but the disc of the larger progenitor remains intact and the newly formed stars are added to the disc. The contribution of disk instability to bulge growth is relatively small in this model.

For galaxies in the semi-analytic output, while it is not straightforward to obtain the photometrically based or kinematically based galaxy types as for the Illustris galaxies, we use the bulge-to-total stellar mass ratio $(B/T)_{\rm SAM}$ to define galaxy morphology. The earlier a galaxy type is, the larger $(B/T)_{\rm SAM}$ it has. Minor mergers make the bulge stellar mass grow gradually, while major mergers can destroy the pre-existing disk completely and form an elliptical galaxy with $(B/T)_{\rm SAM}=1$. To define late and early type galaxies in the semi-analytic model results, we also use a value of 0.6 as the divider, the same as that used for the Illustris galaxies when their morphology types are kinematically determined. As can be seen later in section 3.2 and Fig.~\ref{fig:BTdis}, there are few galaxies with $(B/T)_{\rm SAM}$ in the range of 0.5 to 0.8. Therefore the result presented in the following would remain similar if changing the divider value within this range. 

\section{morphology comparisons}
\label{sec:statistic}

\subsection{Matching galaxies in the two models}

In order to make detailed comparison of galaxy morphology in the two galaxy formation models, we have performed one-to-one matching of the Illustris and the SAM galaxies according to their positions and their parent halo masses. In the SAM model, the position of a galaxy is associated with its parent dark matter halo, while for the full-physics Illustris simulation, due to the impact from baryonic processes, the exact position of the match halo can be different from those in the Illustris-Dark simulation and hence the galaxies in the SAM results. To select the genuine matched haloes/galaxies pairs, we match central galaxies in the two model outputs by requiring that the matched galaxies have similar positions with deviation less than $0.1h^{-1}$Mpc, and that their parent dark matter halo masses differ by less than $50$ per cent. We set this limitation to identify genuine matched pairs, but may not single out all the potential matched galaxies.

In the following, we compare galaxy morphology defined by different methods in the Illustris and the SAM, for the matched galaxy pairs we identified. We focus on galaxy pairs with central stellar masses larger than $10^{10}M_{\odot}$ in the Illustris simulation. The {\it central} stellar mass is taken from the Illustris PMSD catalogue provided by \citet{xu2017}, which is defined as the stellar mass within a radius of $30$kpc from the galaxy center defined as the position of the lowest gravitational potential of its host. This central stellar mass closely represents the mass of the tightly-bound galaxy component, and therefore is used throughout the paper when selecting galaxies by mass. A lower limit of central stellar mass of $10^{10}M_{\odot}$ corresponds to 10,000 stellar particles in the tightly-bound galaxy components. 

In the following analysis, to investigate the dependence of galaxy morphology type on galaxy mass, the matched galaxy sample is further divided into three subsamples. While the stellar masses of the matched galaxy pair have some difference in Illustris and in the SAM model, we divide galaxies into three subsamples according to their central stellar mass in the Illustris simulation:
  (1) the Milky Way-mass galaxies (MW) with central stellar masses in the range of $4-8\times10^{10}M_{\odot}$; (2) less massive galaxies (Less) with central stellar masses of
  $1-4\times10^{10}M_{\odot}$; and (3) more massive galaxies (Massive) with central stellar masses greater than $8\times10^{10}M_{\odot}$.

\begin{table}
\caption{For the matched galaxy pairs in the Illustris and the SAM catalogues,
percentiles of galaxy morphology type in Illustris as defined by \citet{xu2017}
and \citet{genel2015}, and in the SAM model, in the three stellar mass bins considered.}
\begin{center}
\begin{tabular}{ccccccc} \hline
{Mass bin} & \multicolumn{2}{c}{light-based type}  & \multicolumn{2}{c}{kinematics-based type} &  \multicolumn{2}{c}{SAM type} \\
{} & \multicolumn{2}{c}{Xu et al. (2017)}  & \multicolumn{2}{c}{Genel et al. (2015)} &  \multicolumn{2}{c}{this work} \\
\hline
 &   LT     &  ET      &   LT         &  ET      &   LT  &  ET     \\ \hline
Less    &  98.3    & 1.7      &  70.9        &  29.1    & 91.9  &  8.1    \\
MW      &  85.0    & 15.0     &  76.6        &  23.4    & 81.9  &  18.1    \\
Massive &  45.1    & 54.9     &  62.6        &  37.4    & 52.3  &  47.7    \\ \hline
\end{tabular}
\end{center}
\end{table}

\subsection{Morphology comparison}

\begin{figure*}
\bc
\hspace{-0.4cm}
\resizebox{16cm}{!}{\includegraphics{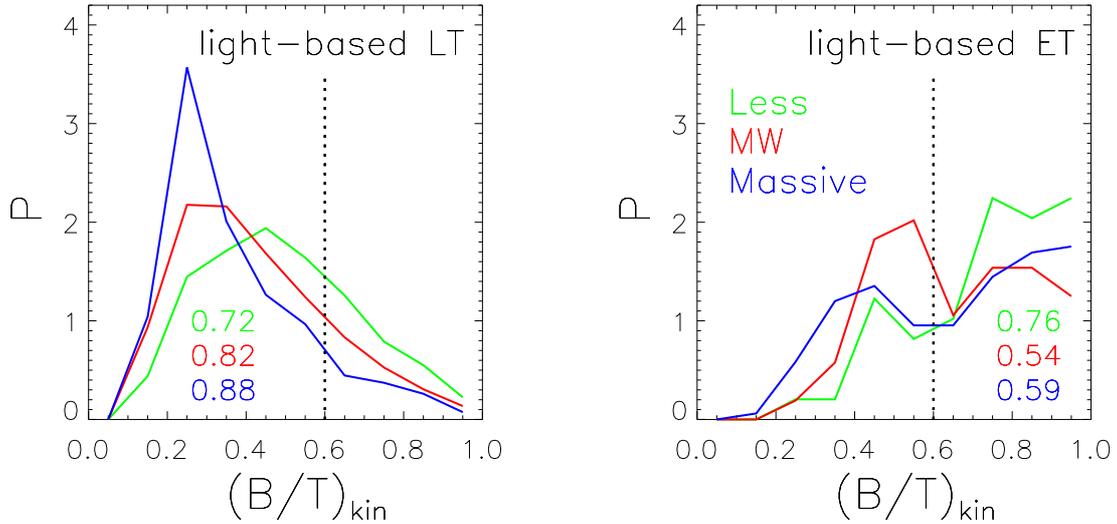}}\\%
\caption{For the matched galaxies in Illustris, the distributions of 
$(B/T)_{\rm kin}$ from \citet{genel2015} for the light-based LT (left panel) 
and ET (right panel) galaxies determined in \citet{xu2017} (normalized
such that the area below the curve is equal to 1). Galaxies in three bins of
central stellar masses are presented: red lines are results for the
Milky Way-mass galaxies with central stellar mass in the range of
$4-8\times10^{10}M_{\odot}$; green lines are for less massive
galaxies ($1-4\times10^{10}M_{\odot}$) while blue lines are for more massive ones
($> 8\times10^{10}M_{\odot}$).
Dotted vertical lines show the value of $(B/T)_{\rm kin}=0.6$, which
is used to define kinematics-based LT and ET galaxies. In the left panel, 
for the light-based LT galaxies, the fractions of kinematics-based
LT galaxies with $(B/T)_{\rm kin}<0.6$ in different mass bins
are shown by numbers in color. In the right panel, for the light-based 
ET galaxies, numbers give the fractions of kinematics-based ET galaxies
in the three mass bins.}
\label{fig:Illustype}
\ec
\end{figure*}

Table 1 lists the fractions of late-type (LT) and early-type (ET) galaxies derived using different methods for the matched galaxies, in the three central stellar mass bins. Disagreements on galaxy types exist to various degrees not only for individual galaxy but also on a statistic level, between the two different methods defining morphology type in the Illustris, and between the hydro-simulation and the SAM results. Below we present more detailed comparisons between results from the light-based and the kinematics-based type determinations of the Illustris hydro-simulation itself (Section 3.2.1) and those between results of Illustris and of the SAM output (Section 3.2.2).

\subsubsection{Comparison between the light-based and the kinematics-based type determinations}
Table 1 shows that galaxy morphology is not exactly consistent between the light-based and kinematics-based type determinations for the Illustris galaxies, and the degree of inconsistence seems to depend on galaxy stellar mass. For galaxies in the ``Less'' and ``MW'' mass bins, the fractions of light-based LT (ET) galaxies are significantly higher (lower) than those of the kinematics-based counterparts. Such findings are consistent with the recent study of \citet[][Figure 7]{bottrell2017}, which found that at lower masses, the photometrically defined bulge fraction is systematically lower than the one defined using kinematic decomposition. 

For galaxies in the ``Massive'' bin, the fraction of LT galaxies according to the light-based type definition is lower than that using kinematics-based type definition.  A good fraction of galaxies that still maintain strong rotation features (and thus less random motion in bulge) have also established de Voucauleurs light distribution. 

Fig.~\ref{fig:Illustype} presents, for the matched galaxies in the Illustris, 
the probability distributions of $(B/T)_{\rm kin}$ from \citet{genel2015} for the 
light-based LT (left panel) and ET (right panel) galaxies as determined by \citet{xu2017}. Green, red and blue lines are results for the ``Less'', ``MW'' and ``Massive'' samples, respectively. In all stellar mass bins, there are galaxies that are light-based LT, with exponential light profile, but are kinematically random motion dominated. There are also galaxies in all stellar mass bins that are light-based ET, preferring a de Voucauleurs light profile, but are kinematically dominated by co-rotating motions.

Nevertheless, in general the type determinations are largely consistent between the two methods. For galaxies in the ``Less'', ``MW'' and ``Massive'' stellar mass bins, the fraction of galaxies with the same morphology types is 72, 82 and 88 for the light-based LT galaxies, respectively. For the light-based ET, the fractions are are 76, 54 and 59 percents, respectively.

\subsubsection{Comparisons between the hydro-simulation and the SAM outputs}

\begin{figure*}
\bc
\hspace{-0.4cm}
\resizebox{11cm}{!}{\includegraphics{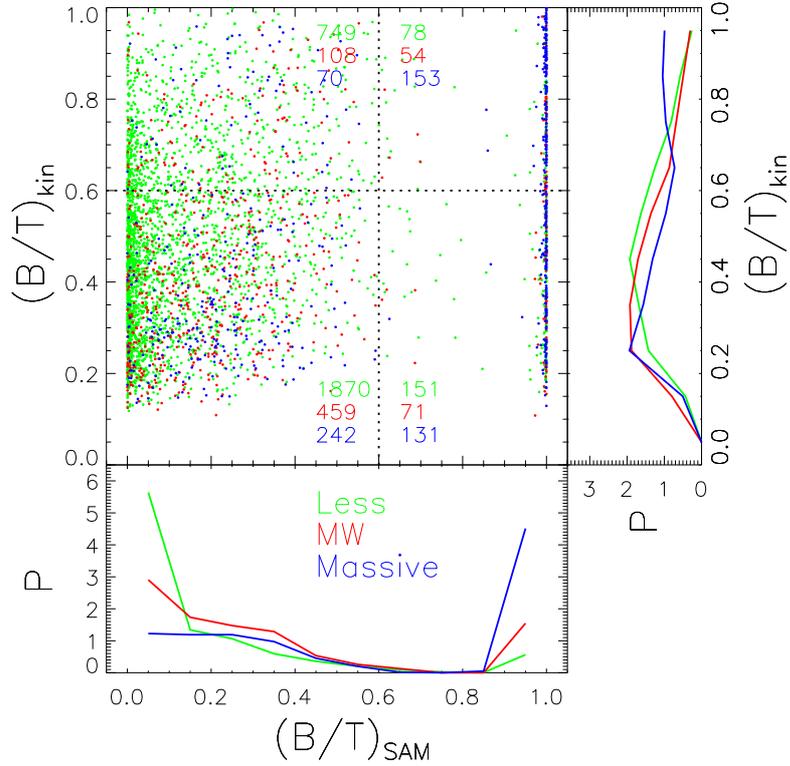}}\\%
\caption{ 
B/T comparison between Illustris and the SAM catalogue, for the matched galaxy 
pairs, and in three stellar mass bins indicated by different color. 
Upper left panel: one-to-one comparison between the SAM $(B/T)_{SAM}$ 
and the Illustris kinematics-based $(B/T)_{kin}$. Vertical and horizontal
dotted lines mark the value of $B/T=0.6$, which is used to separate LT and ET
galaxies in the two methods. Numbers of galaxies included in each of the
four rectangle regions divided by the two dotted lines are presented, 
in the three mass bins shown by different colors. 
Bottom panel: the 
distributions of $(B/T)_{SAM}$ (normalized such that the area below the 
curve is equal to 1). Right panel: the distributions of $(B/T)_{kin}$ 
for Illustris galaxies in different stellar mass bins.}
\label{fig:BTdis}
\ec
\end{figure*}

Fig.~\ref{fig:BTdis} shows the comparison of bulge-to-total ratios in the SAM model and in the Illustris provided by \citet{genel2015}. The upper left panel presents the one-to-one
comparison for all matched galaxies in the three mass bins. The bottom and the right panel display the distributions of $B/T$ of the SAM result and of the Illustris result, respectively. As can be seen, the $B/T$ distributions are clearly different between the two: the kinematic ratios $(B/T)_{kin}$ of the hydro-simulation galaxies follow a broad distribution between 0 and 1, while the $(B/T)_{SAM}$ distribution shows a much more prominent bimodality with peaks around 0. In addition, there are very few galaxies with $B/T$ between $0.5\sim0.8$ for the Illustris galaxies.

This indicates that galaxy morphology type transformation takes place much more
gradually in the hydrodynamic simulation than in the SAM. This is because of the treatment of elliptical galaxy formation in SAM where galaxies turn their morphologies from disk- to bulge-dominant systems immediately after a major merger. In both cases, massive galaxies have a larger fraction of $B/T~>0.8$ than low-mass galaxies. For the SAM result, the fraction of galaxies with $B/T<0.2$ also strongly depends on galaxy masses, with less massive galaxies having higher fraction of low $B/T$.

\begin{table}
\caption{
For Illustris galaxies with a light-based or a
kinematics-based morphology type, the percentile fractions of the
matched galaxies in SAM that have a consistent type determination
in different stellar mass bins. }
\begin{center}
\begin{tabular}{cccc} \hline
 Type & Mass bin &  light-based &  kinematics-based  \\ \hline
  & & Xu et al. (2017) & Genel et al. (2015)  \\ \hline
       LT & Less     &  92.2    & 92.5    \\
          & MW       &  84.4    & 86.6    \\
          & Massive  &  70.3    & 64.9    \\ \hline
       ET & Less     &  20.4    & 9.4     \\
          & MW       &  31.7    & 33.3    \\
          & Massive  &  62.4    & 68.6    \\ \hline
\end{tabular}
\end{center}
\end{table}

In the upper left panel of Fig.~\ref{fig:BTdis}, vertical and horizontal dotted lines mark the value of $B/T=0.6$, which separate LT and ET galaxies in the two models. In each of the four rectangle regions divided by the two dotted lines, numbers in different colors are the numbers of galaxies included in each region in three mass bins. For example, for the MW mass bin, the number of galaxies that are kinematics-based LT in Illustris, while LT(ET) in SAM is 459(71) 
shown in red. Therefore, the bottom left rectangle includes galaxies that are consistently LT, while the upper right rectangle includes galaxies that are both ET in the two models. Galaxies in the other two regions have inconsistent types. In the following, we will show in more detail for Illustris galaxies of a given type, the distributions of $(B/T)_{SAM}$ of the matched galaxies, and list the fractions of matched galaxies that have consistent type determination in SAM in different mass bins.

\begin{figure*}
\bc
\hspace{-0.4cm}
\resizebox{15cm}{!}{\includegraphics{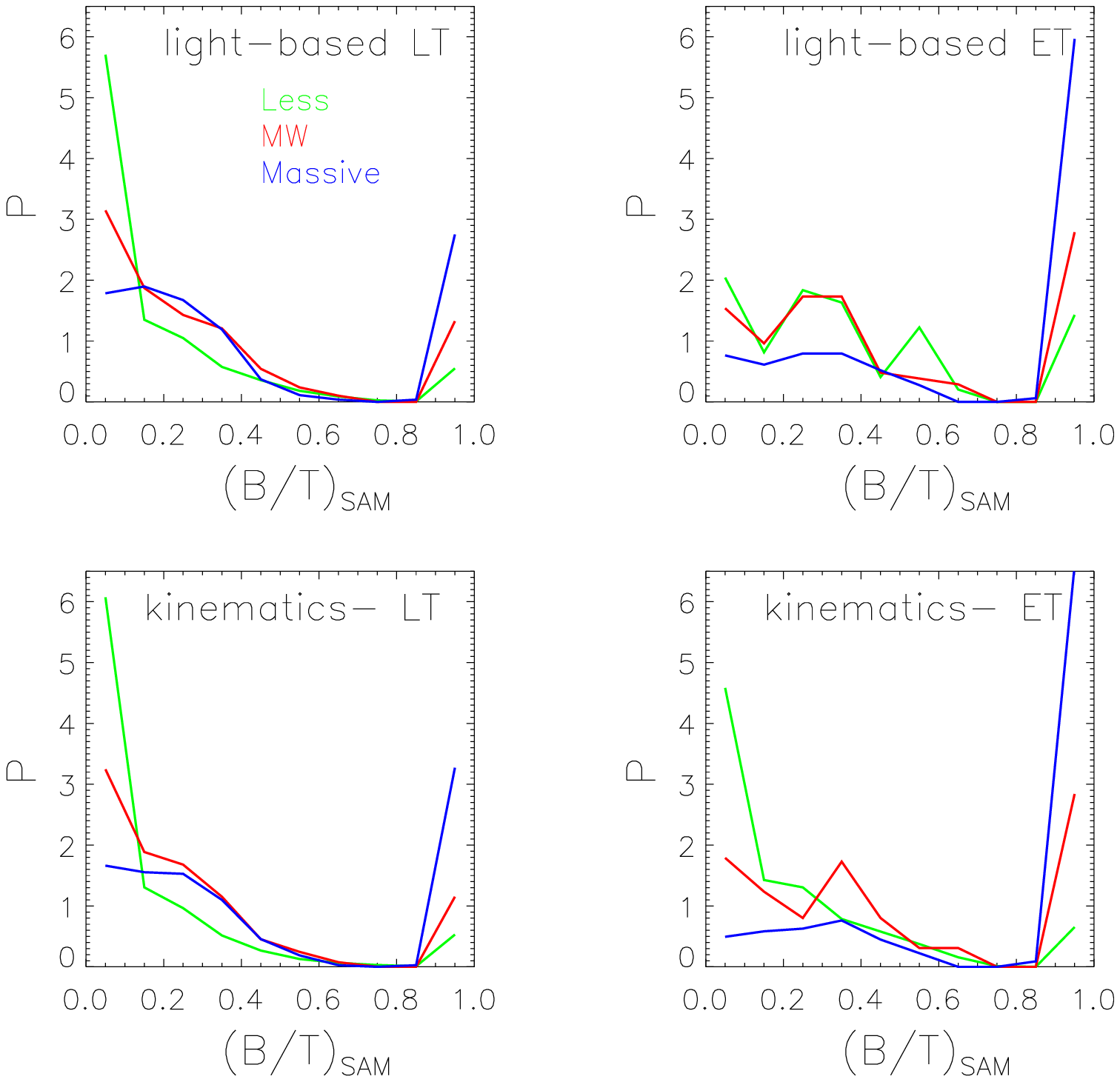}}\\%
\caption{$(B/T)_{\rm SAM}$ distributions of the matched SAM galaxies, for LT (left panels)
and ET (right panels) galaxies in Illustris, and in three mass
intervals  (normalized such that the area below the curve is equal to 1). 
Upper panels are the results when morphology types in
Illustris are defined light-based by \citet{xu2017}, and lower panels are for
types defined kinematics-based by \citet{genel2015}.  }
\label{fig:matchcompare}
\ec
\end{figure*}

In Fig.~\ref{fig:matchcompare}, for the Illustris galaxies of a given morphology 
type,  we plot the distributions of $(B/T)_{\rm SAM}$ of their SAM counterparts. 
Upper panels show the results for the light-based Illustris LT (upper left panel) and ET galaxies (upper right panel). Lower panels are the results when the types of Illustris galaxies are determined kinematically. Although as shown in the section 3.2.1, the morphology types of the Illustris galaxies are not exactly consistent when defined with light-based and kinematics-based methods, the overall distributions of $(B/T)_{\rm SAM}$ shown in Fig.~\ref{fig:matchcompare} are very similar, suggesting that above results are largely independent on the definition of the morphology type in the Illustris simulation. 

For the Illustris LT galaxies, most of them are of the same types in the SAM output
with $(B/T)_{\rm SAM} < 0.6$. For the Illustris ET galaxies, however, the degree of consistency is much lower, a large fractions of their SAM counterparts have different types, especially for the ``Less'' and ``MW'' samples.
A more quantitative comparison is given in Table 2. For Illustris galaxies with a given type determination defined light-based/ kinematics-based, Table 2 lists the percentile fractions of the matched galaxies in SAM that have consistent type determination as the Illustris one according to their $(B/T)_{SAM}$, in different stellar mass bins. 

From the numbers in Table 2 we see more clearly the consistency between the two models depends on stellar mass.  The consistency of LT galaxies is very high for lower-mass galaxies, and decreases to about 70 per cent in the most massive bin. For ET galaxies the consistency depends much strongly on galaxy stellar mass. Very small fraction of SAM galaxies are also ET in the lowest mass bin. In the most massive bin, the consistency is much better, and reaches to more than 60 per cent.

\citet{rodriguez2017} studied recently the importance of mergers and halo spin in shaping galaxy morphology in the Illustris hydrodynamical simulation. They found that mergers play a dominant role in shaping the morphology of galaxies more massive than $10^{11}M_{\odot}$, while the morphology of the Milky Way-mass galaxies shows little dependence on galaxy assembly history or halo spin. Since mergers basically
determine the morphology of galaxies in the SAM model, this is consistent with our result that most massive galaxies have similar morphology in the two methods, while for Milky Way-mass galaxies the consistency of early type galaxies in the two methods is quite low.

\section{Case study: comparison of galaxy growth history}
\label{sec:example}

In order to investigate why some galaxy morphology types predicted by the Illustris and the SAM are inconsistent, we have compared the detailed growth and merging histories of some Milky Way-mass galaxies in the two models. We show below one of them as an example and discuss our findings.

\begin{figure*}
\bc
\hspace{-0.4cm}
\resizebox{15.5cm}{!}{\includegraphics{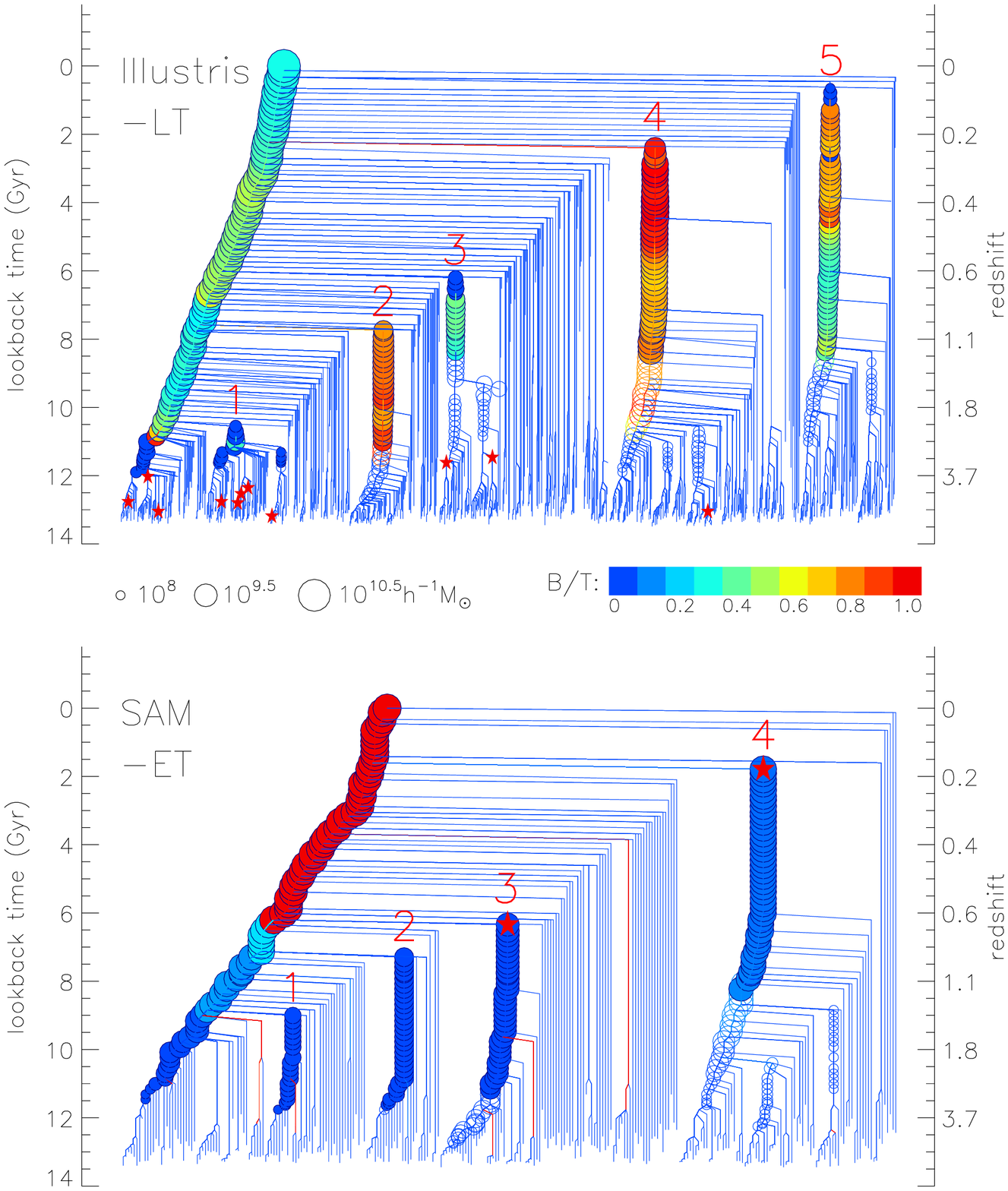}}\\%
\caption{ 
The merging trees for an example matched galaxy pair that is LT in Illustris (upper panel)
but ET in the SAM model (lower panel). Galaxies with total stellar mass greater than $10^8M_{\odot}$ are shown by circles with size proportional to $log(M_{stars})$, while less massive ones are shown by dots. The most left symbols give the main branch of the progenitors. Color of each circle shows the B/T ratio of the galaxy, using $(B/T)_{\rm kin}$ for Illustris galaxies in the upper panel and for $(B/T)_{\rm kin}$ for SAM galaxies in the lower panel. Progenitor and the descendant galaxies are linked by lines with the same color as the progenitor.  Filled circles are progenitors that are already in the main FOF group of the main branch progenitors, while open circles are the ones that are not yet into the group. Red stars mark the major merger events where the ratio between stellar mass of the two merged galaxies is greater than $0.3$, on the galaxies with smaller stellar mass in the major mergers. Small red stars mark the major mergers where both merged galaxies have stellar mass less than $10^8M_{\odot}$. Large red stars are for the major mergers where at least one merged galaxy has stellar mass greater than $10^8M_{\odot}$. Sub-branches marked with the same numbers have one-to-one correspondence in the upper and lower panels.
Sub-branch 5 in the upper panel in Illustris corresponds to an ``orphan''  galaxy in the SAM model, and therefore is not included in the merger tree of the central galaxy in the lower panel.}
\label{fig:example_LT}
\ec
\end{figure*}



Fig.~\ref{fig:example_LT} shows the merging trees of the galaxy, which is LT in Illustris (upper panel) but ET in the SAM model (lower panel). Relatively bigger branches are labeled with numbers in the figure, and these numbers in the upper and lower panels have one-to-one correspondence between the two models. 
Sub-branch 5 in the upper panel in Illustris corresponds to an ``orphan''  galaxy in the SAM model and is therefore not included in the merger tree of the central galaxy (the explanations of this point and other intrinsic differences between the merger trees of the two models can be found in the Appendix). 

In the SAM model, the main-branch galaxy is a pure disk galaxy at early times. At lookback time of around 9 Gyr, a merger with a satellite galaxy (sub-branch 1) happens (with the stellar mass ratio of the merging galaxies to be 0.15) and builds a small bulge component. A subsequent prominent merger with sub-branch 2 (merger ratio to be 0.18) continues to build the bulge component obviously. At lookback time of around 6.3Gyr, a major merger with sub-branch 3 (merger ratio to be 0.40) turns the galaxy into an elliptical with B/T=1. Since then, the galaxy remains as early-type and does not build a significant disk component again. 

In the hydrosimulation, the matched galaxy has similar merger history as in the SAM model in general but not in details. For examples, the exact merging time, the mass ratio of the merging galaxies and the impact of mergers on the resulting galaxy morphology differ. Initially the main-branch galaxy is a late-type galaxy and almost retains its morphology type all the way to redshift 0. Recall that because of the major merger with sub-branch 3, its counterpart in the SAM turns into an early-type. However this does not happen to the matched galaxy in Illustris. Interestingly, the merger with sub-branch 3 in Illustris has a mass ratio of 0.02, much smaller than that in the SAM. This merger consequently only results in a minor merger at the end and does not change the galaxy morphology type. Indeed, mergers always have lower mass ratios in Illustris when comparing with the SAM and so have smaller effect on changing morphology, which can be clearly seen in Fig.~\ref{fig:example_LT}.


In summary, by comparing the merger histories of this example galaxy, we see that by construction, galaxy morphology in the SAM model is closely related to mergers: the higher the merger mass-ratio is, the more prominent the formed galactic bulge is. When major mergers happen, the galaxy turns into early type with $B/T=1$. In the hydrosimulation, however, galaxy morphology does not correlate with mergers as tightly as in the SAM model. Even with several prominent mergers as shown above, the galaxy can remain as late-type in Illustris. We have also examined and compared the merger histories of some other galaxies in the two models. We find that in the Illustris hydrodynamical simulation, even when major mergers happen, galaxies may not necessarily become bulge-dominant.  The same fact has been found in many previous studies of hydro-simulations \citep[e.g. ][]{barnes2002, springel2005c, governato2009, athanassoula2016, eliche2018}, although most simulations still produce too massive bulges compared with real spirals \citep{brooks2016, athanassoula2016}. On the other hand, galaxies can grow significantly without strong mergers in Illustris, similar as the galaxies in sub-branch 2 and 4 in Fig.~\ref{fig:example_LT}.

Another clear difference between the two models lies on the mass of the satellite galaxies. In Illustris, the mass of a satellite decreases as it approaches/falls into the central galaxy \footnote{ The dramatic decreasing of satellite stellar mass is partially due to the assignment of mass in the halo finding method adopted in Illustris, where most of the loosely bound matter residing in the FoF group is assigned to the central galaxy rather than the satellite, as shown and discussed in Figure 3 of \citet{rodriguez2015}.}, due to ram pressure and/or tidal stripping.
In the SAM, however, the satellite stellar mass would not decrease, since the stellar component is assumed to be at the innermost part of the galaxy and to be largely unaffected when the outer dark matter/gas components get completely stripped. 
As a result, Illustris has mergers with smaller mass ratios and with satellites having more loosely bound mass. Therefore in general less severe mergers are seen in Illustris than in the SAM model.


\section{discussion and conclusion}
\label{sec:conclusion}

In this work, we compare galaxy morphology predicted by the hydrodynamical simulation Illustris and a semi-analytic model combined with the Illustris-Dark simulation. Morphologies of Illustris galaxies are determined based on the light profile as well as the kinematic feature of galaxies, and morphologies of the semi-analytic model galaxies are defined according to the bulge-to-total stellar mass ratio. Our findings can be summarized as follows, and are largely independent on the two different definitions of the morphology type in Illustris.

Illustris galaxies change their morphologies from late-type to early-type gradually during their evolution, with a wide and extended distribution of bulge-to-total mass ratio. For galaxies in the SAM model, by definition, the formation of galactic bulges relates closely to mergers, and results in a distribution of bulge-to-total mass ratio  concentrating on the low and very high mass ends.

For late-type galaxies in the hydro-simulation, more than $80\%$ of their counterparts in the SAM model have similar types at Milky Way-mass scale and below; the fraction drops to $\sim 65-70\%$ for more massive galaxies. The consistency in galaxy morphology between the SAM and hydro-types are much poorer among early-type galaxies. In particular, for early-type galaxies in the hydro-simulation, the fractions of SAM counterparts that are also early-types are remarkably low, which are less than $\sim 34\%$ for galaxies with mass comparable to or less than the Milky Way. For more massive galaxies, the consistency rate rises to more than $60\%$.

In order to understand the causes of the difference between hydro-simulation
and the SAM model, we have compared in detail the merger histories of some matched galaxies in the two models and show one example galaxy with inconsistent morphology type. Two aspects of differences between the hydro- and SAM morphologies are notable. Firstly, in SAM models, major mergers and frequent minor mergers would result in the growth of bulges and turn galaxies into early-types in a predominant fashion. In hydro-simulations, however, bulge formation is not always as tightly connected to mergers as it is in SAM: bulges can grow prominently without strong merger events, and mergers with massive satellites do not necessarily result in early-type morphology. Secondly, the mass ratios of mergers in Illustris are much smaller than those in SAM because of different treatment of tidal stripping of stellar component of satellite galaxy. 

While we focus on comparing only galaxy morphology in a hydro-simulation and a semi-analytic model, studies such as \citet{guo2016} and \citet{mitchell2017} have done more detailed comparison between the two methods. They both focused on galaxy size rather than late/early type morphology as we do. \citet{guo2016} compared various galaxy properties predicted by hydro-simulation and by two SAM models, and showed that different SAM models predict a bit different stellar mass - sizes relations. 
\citet{mitchell2017} gave a comprehensive analysis on the different physics adopted in  typical hydro-simulation and SAM model. Note that in most SAM models as the one analysed in this study, the formation of bulge component is mainly determined by the treatment of merger-induced bulge growth, although in detail how massive the bulge can be is also affected by the physics assumed. In hydro-simulations, the bulge growth and general morphology change should be more closely related to how the physics are treated. \citet{fontanot2015} have tested that when implementations similar as hydro-simulation are applied, i.e., when including baryonic mass transfer between different components during mergers, the efficiency of bulge formation through mergers would be reduced in SAM. 

With advanced numerical methods, how galaxies shape their morphology of present day has been studied in various cosmological hydro-simulations \citep{thob2018, garrison2018, rodriguez2019}, although tensions still remain in producing for example, the observed galaxy size and the morphology-colour relation \citep{rodriguez2019}. \citet{trayford2019} studied galaxy morphology transition since z=1 using EAGLE simulations, and found that for galaxies more massive than $10^9M_{\odot}$, 60 per cent of the disc-to-spheroid morphology transitions are not due to mergers. This is qualitatively consistent with our result for Illustris that merger and morphology transition are not tightly correlated in hydro-simulations.
Nevertheless, some questions remain to be answered. For instance, it is not clear yet why in hydrodynamical simulations mergers sometimes result in bulge-dominant morphology and sometimes not. We will investigate these issues in detail in our future work.

\section*{Acknowledgments}
We thank Volker Springel for helpful comments, and Dylan Nelson for providing videos of the merging histories of galaxies in Illustris to help us understand the processes better. LW acknowledges support from the NSFC grants program (No.11573031), and the National Key Program for Science and Technology Research and Development (2017YFB0203300). QG is supported by two NSFC grants (Nos. 11573033, 11622325).

\bibliographystyle{mnras}
\bibliography{morphology}

\appendix
\section{Intrinsic differences between galaxy merging trees in the two models}

When comparing merging trees for the matched galaxy pairs in Illustris and in the SAM model, we note that there are more first progenitors/galaxies (the galaxies that have no progenitors) and more small branches in the merger tree of the galaxy in the Illustris than in the SAM model, as shown in Fig.~\ref{fig:example_LT}. Three possible reasons could account for this.  

Firstly, compared to hydrodyamical simulations, SAM galaxy merging trees may miss some first galaxies due to finite temporal resolution recording the evolution of haloes/subhaloes. In the SAM model we use, the stellar component is initialized only in central halo when the halo is able to make stars at a recorded time/snapshot, therefore neglecting star formation in some haloes who may have grown and made stars between two adjacent snapshots but have become subhaloes at the recorded snapshot. The Illustris galaxy merging history does not suffer from this issue. 

Secondly, the SAM model comprises a population of ``orphan'' galaxies \citep{wang2006} whose host subhaloes are tidally disrupted due to finite numerical resolution. These ``orphan'' galaxies have their own trees and thus are not included in the merger tree of this central galaxy at z=0 in the SAM result. While the Illustris simulation does not have the ``orphan''  galaxy population,  the numerically ``overmerging'' galaxies are included in the merger tree. An example of ``orphan''  galaxy not included in the merger tree of the central galaxy in the SAM is seen in Fig.~\ref{fig:example_LT} (sub-branch 5 in the upper panel).

Apart from the two main points above, due to intrinsic difference of the two simulations, and also the different SubLink and LHaloTree trees used as described in section 2.1, there also exists some differences between the Illustris and the Illustris-Dark (sub)halo merger trees, of which the latter is the input of the SAM model.

\bsp	
\label{lastpage}
\end{document}